\documentclass[letterpaper, 10 pt, conference]{ieeeconf}  % Comment this line out if you need a4paper

\IEEEoverridecommandlockouts                              % This command is only needed if 
\usepackage{graphicx}
\usepackage{csquotes}
\usepackage{subfigure}
\usepackage{multirow}
\usepackage{graphics} % for pdf, bitmapped graphics files
\usepackage{epsfig} % for postscript graphics files
\usepackage{mathptmx} % assumes new font selection scheme installed
\usepackage{times} % assumes new font selection scheme installed
\usepackage{amsmath} % assumes amsmath package installed
\usepackage{amssymb}% assumes amsmath package installed
\usepackage{cite}
\title{\LARGE \bf
Comparative Analysis and Calibration of Low Cost Resistive and Capacitive Soil Moisture Sensor
}

\author{Sourodip Chowdhury$^{*}$, Shaunak Sen$^{*}$ and S Janardhanan$^{*}$% <-this % stops a space
\thanks{$^{*}$Department of Electrical Engineering, Indian Institute of Technology, Delhi-110016.}
}

\begin{document}

\maketitle
\thispagestyle{empty}
\pagestyle{empty}

%%%%%%%%%%%%%%%%%%%%%%%%%%%%%%%%%%%%%%%%%%%%%%%%%%%%%%%%%%%%%%%%%%%%%%%%%%%%%%%%
\begin{abstract}

Soil moisture is an essential parameter in agriculture. It determines several environmental and agricultural activities such as climate change, drought prediction, irrigation, etc. Smart irrigation management requires continuous soil moisture monitoring to reduce unnecessary water usage. In recent times, the use of low-cost sensors is becoming popular among farmers for soil moisture monitoring. In this paper, a comparison of low-cost resistive and capacitive soil moisture sensors is demonstrated in two ways. One way is the calibration of the sensors in gravimetric and volumetric water content, and the other is the sensors' response analysis when different quantities of water are added to the same amount of soil. The analysis shown in this work is essential before choosing cost-effective sensors for any soil moisture monitoring platform.

\end{abstract}

%%%%%%%%%%%%%%%%%%%%%%%%%%%%%%%%%%%%%%%%%%%%%%%%%%%%%%%%%%%%%%%%%%%%%%%%%%%%%%%%
\section{INTRODUCTION}
Soil moisture (SM) is an essential agricultural parameter used to estimate various environmental and agricultural activities such as climate change, drought prediction, and irrigation scheduling, especially for estimating water stress in agrarian land~\cite{GRUBER2022,wicki2020assessing}. Thus it has become the center of attraction among other agricultural research areas. Adequate prediction and estimation of different environmental and climatic variables require in-situ measurement to be reliable and precise~\cite{berg2018climate}.
\par SM measurement can be carried out in two ways, direct measurement and indirect measurement. The gravimetric method is the direct and natural way to measure SM. However, it is very labor-intensive and time-consuming, although it accurately gives ground truth data of soil moisture content~\cite{mukhlisin2021techniques,r1}. All other SM measurement technique comes under the indirect measurement method, indirect because they use the change in the internal property of the sensory system as a proxy for soil water content (SWC). Apart from the gravimetric method, ground contact type sensors can measure direct information on soil moisture content (SMC). In the ground contact type sensor category, time-domain reflectometry, time-domain transmission, frequency domain reflectometry, neutron probe, electrical resistance, electromagnetic sensors, and tensiometers are widely used ~\cite{r1,john2021temperature,zhu2019time,pereira2020automation,singh2019soil}. However, due to the setup cost and bulkiness, these methods are less preferred for soil moisture monitoring~\cite{r1,r6,sharma2018methods,mukhlisin2021techniques}.
\par SM monitoring using the Internet of Things (IoT) based intelligent sensor system is prominently used in the recent era~\cite{r3}. However, before developing an IoT-based SM monitoring, we need to focus on the sensor performance. Predominantly capacitive and resistive sensors are used for IoT-based SM monitoring system~\cite{nagahage2019calibration,adla2020laboratory} as the sensors are less costlier and have easy setup requirements than other ground contact sensors. The ground contact sensors are an indirect method of SM measurement, so they need calibration. The resistive sensor works based on the change in conductivity, and the capacitive sensor works based on the change in the dielectric constant. They provide output in terms of voltage. So they need to be calibrated in terms of the SWC~\cite{nagahage2019calibration,adla2020laboratory,hrisko2020capacitive} before integrating into a moisture monitoring platform. 
\par Calibration of the low-cost capacitive sensor has been performed  in~\cite{GRUBER2022,nagahage2019calibration} and validation with the gravimetric method shown for an automated soil moisture monitoring platform. Various investigation on the calibration method, calibration performance, and verification has been conducted in~\cite{adla2020laboratory,muzdrikah2018calibration,kizito2008frequency}. A comparison of low-cost resistive and capacitive sensors was investigated in~\cite{adla2020laboratory,songara2022calibration} and the calibration performance is compared using $R^2$ value, RMSE, MAPE, and RAE.

\par In work, we aim to investigate the performance of resistive and capacitive sensors based on different features to compare their applicability and reliability. Initially, we compare the calibration results based on the $R^2$ value, RMSE, and the variance on repetition. Later they are compared based on the response for a different amount of water added to the same soil.

\section{Materials and Methods}
\subsection{Soil Moisture Sensor}
Soil moisture is the interpretation of liquid or water content in the soil. There are many methods used for soil moisture measurement. Direct soil moisture measurement is done using the gravimetric method, but due to the rigorousness and labor intensiveness, it is not preferred for monitoring soil moisture. Indirect methods are based on the measurement of the change in physical properties of the soil as a proxy for soil moisture. Sensors generally work on the indirect principles, where the difference in the physical quantity reflects an electrical quantity.
\par The capacitive soil moisture sensor works on the principle of change in dielectric constant due to change in SWC. The capacitance-based sensing probe comes within a module that includes 555 timer IC to count the resonant frequency as an indirect measure of SWC. The resistive soil moisture sensor operates on the idea of change in resistance when current conductivity between the probes changes in response to a change in SWC. The change in the moisture content is measured using the LM393 comparator circuit that comes with the sensor probes. The raw output from both the sensors is in terms of voltage, digitized in the range 1 to 1024. The raw data or the output voltage is inversely proportional to the SWC.
\begin{equation}
    \text{Voltage Output} = \frac{\text{Raw Sensor Output}}{1024} \times 3.3\ V
    \label{equ: 1}
\end{equation}
\subsection{Data Acquisition System}
A NodeMCU and a power supply unit are included in the data acquisition system. NodeMCU is an open-source firmware and development board with an integrated ESP-8266 WiFi module that allows the board to connect to the internet and send sensor data to the cloud.
\subsection{Cloud Platform}
Thingspeak~\cite{nagahage2019calibration} is an open-source IoT analytics platform by Mathworks. It is generally used for storing, visualizing, and monitoring sensor data. Thingspeak provides different channels and corresponding read and writes API keys, using which data are sent to the platform. Data can be read directly from the Thingspeak medium using the read API key through MATLAB online for further analysis.
\subsection{Soil Water Content (SWC)}
SWC is the measure of soil moisture, it is interpreted in two ways, gravimetric water content (GWC, denoted as $\theta_g$) and volumetric water content (VWC, denoted as $\theta_v$)~\cite{hrisko2020capacitive,al2018soil}. GWC is measured as the weighted amount of water present per weighted amount of soil. VWC is measured as the volume of water present per volume of soil. The interrelation between GWC and VWC is given below,
\begin{equation}
    \centering
    \theta_g = \frac{M_{w}}{M_{s}} = \frac{M_{wet} - M_{d}}{M_{d}}.
    \label{equ: 2}
\end{equation}
$M_{w}$, $M_{s}$, $M_{wet}$ and $M_{d}$ are the mass of the water present in the soil, mass of the soil, mass of the wet soil and mass of the dry soil respectively.
\begin{equation}
    \centering
    \theta_v = \frac{V_{w}}{V_{s}} = \frac{\frac{M_{w}}{\rho_w}}{\frac{M_s}{\rho_s}},
    \label{equ: 3}
\end{equation}
\begin{equation}
    \centering
    \theta_v = \frac{M_w}{M_s} \cdot \frac{\rho_s}{\rho_w},
    \label{equ: 4}
\end{equation}
\begin{equation}
    \centering
    \theta_v = \theta_g \cdot \rho_s.
    \label{equ: 5}
\end{equation}
$\rho_w$ is the density of water, assumed to be 1 $gm/cm^3$. Thus the final equation \ref{equ: 5} shows a relation between GWC and VWC, where $\rho_{s}$ is the bulk density of the soil. The bulk density of soil is calculated as the ratio of the weight of dry soil to the volume of the soil, given in Equation~\ref{equ: 6}~\cite{al2018soil}.
\begin{equation}
    \centering
    \rho_s = \rho_{bulk} = \frac{\text{Weight of dry soil}}{\text{Volume of soil}}.
    \label{equ: 6}
\end{equation}
The dry weight of soil is calculated using the GWC from the equation \ref{equ: 2},
\begin{equation}
    \centering
    \text{Dry soil weight} = \frac{\text{Weight of the soil (wet)}}{1 +\text{GWC}}.
    \label{equ: 7}
\end{equation}

\subsection{Calibration Method}
 The experiment has been performed in Control Laboratory in the Department of Electrical Engineering at IIT Delhi. The calibration is performed using clay type soil available in the campus.
\begin{figure}[ht]
    \centering
    \includegraphics[width = 8cm]{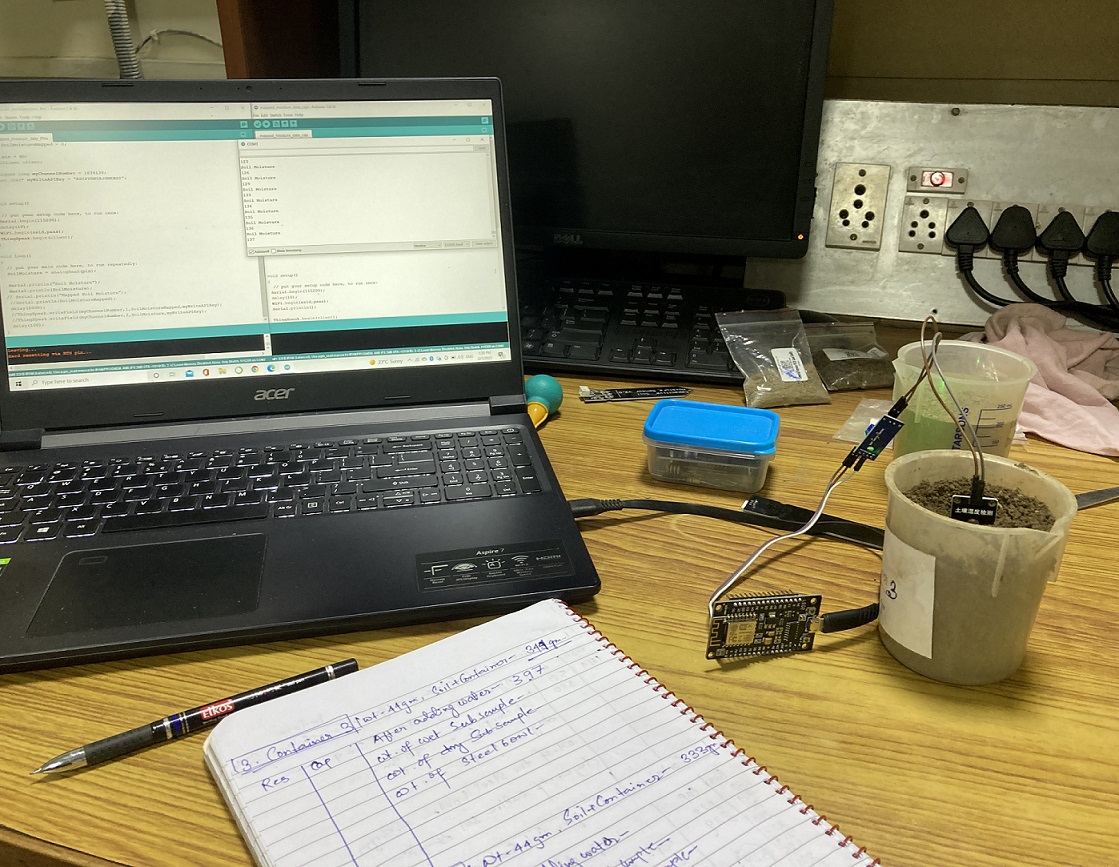}
    \caption{Sensor Data collection from different soil sample}
    \label{fig:cal_data}
\end{figure} 

\par
\begin{figure}[th]
    \centering
    \includegraphics[width =8.5cm]{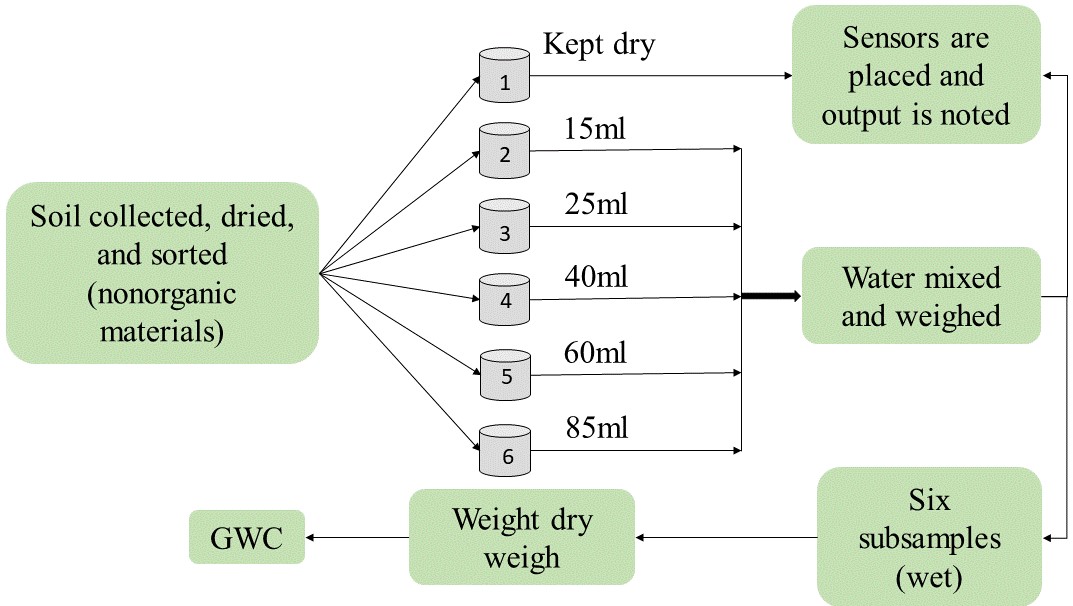}
    \caption{Experimental Design}
    \label{fig:cal_flow}
\end{figure}

 The experimental requirements were six containers, one oven for drying purpose, sensor arrangements, zipped lock plastic bags, weighing machine with $10mg$ resolution. The calibration experiment flow is given in Fig.~\ref{fig:cal_flow}. The experimental outcome produce the GWC value, the VWC is then calculated using the obtained GWC and bulk density of each soil volume.
\subsection{Sensor Response Study}
The response study of the resistive and capacitive soil moisture sensors has been performed for different quantities of water addition to the same quantity of soil. This experiment is planned in such a way that the sensor response is independent of any other environmental parameters, such as ambient temperature variation, humidity to compare the results. The experiment has been performed in laboratory and the experiment data is sent to Thingspeak, later they are collected and processed for drawing insights.

\section{Results and Data}

\subsection{Calibration}
\begin{figure}[th]
    \centering
    \subfigure[Capacitive Sensor]
    {
        \includegraphics[width = 4cm]{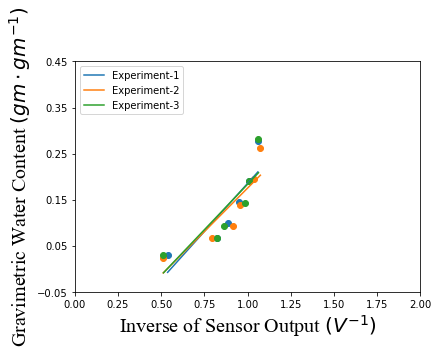}
        \label{fig:cal_gwc}
    }
    \subfigure[Capacitive Sensor]
    {
        \includegraphics[width = 4cm]{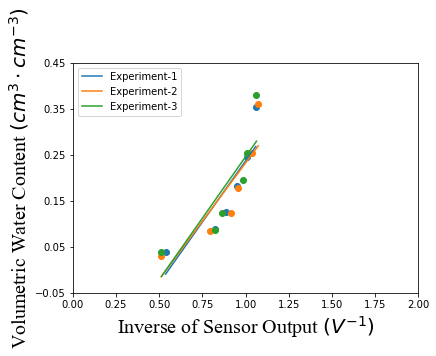}
        \label{fig:cal_vwc}
    }
    \subfigure[Resistive Sensor]
    {
        \includegraphics[width=4cm]{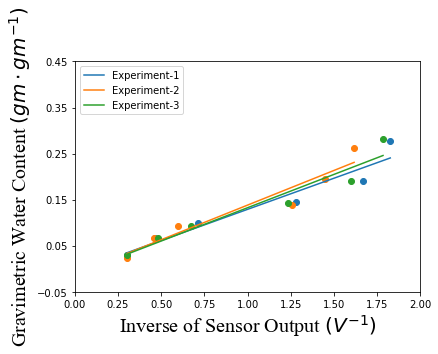}
        \label{fig:res_gwc}
    }
    \subfigure[Resistive Sensor]{
        \includegraphics[width = 4cm]{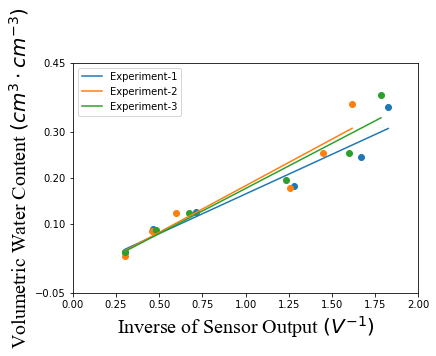}
        \label{fig: res_vwc}
    }
    \caption{Calibration for soil moisture sensors}
    \label{fig:Calibration}
\end{figure}

% \begin{figure}
%     \centering
%   \subfloat[a\label{1a}]{\includegraphics[width=0.42\linewidth]{cap_cal_fit_gwc.png}}
%   \subfloat[b\label{1b}]{\includegraphics[width=0.42\linewidth]{cap_cal_fit_vwc.png}}
% %     \\
% %   \subfloat[c\label{1c}]{%
% %         \includegraphics[width=0.45\linewidth]{example-image}}
% %     \hfill
% %   \subfloat[d\label{1d}]{%
% %         \includegraphics[width=0.45\linewidth]{example-image}}
%   \caption{(a), (b) Some examples from CIFAR-10 \cite{4}. The objects in     
%         single-label images are usually roughly aligned.(c),(d) However, the 
%         assumption of object alignment is not valid for multi-label
%         images. Also note the partial visibility and occlusion
%         between objects in the multi-label images.}
%   \label{fig1} 
% \end{figure}
Laboratory-based soil moisture sensor calibration is performed, and the data for six different soil samples are collected using the sensor-based data acquisition system. Soil samples are made by mixing an increasing array of water till the sixth container saturates. This experiment has a total of three replications. All the experimental data are plotted as shown in Fig.~\ref{fig:Calibration}. The sensor output is inversely proportional to the SWC. Thus the inverse of the sensor (voltage) output is plotted against SWC. SWC is calculated in two ways, GWC and VWC, where the GWC is calculated using Equation~\ref{equ: 2} and VWC is calculated using Equation~\ref{equ: 5}.

 The Experimental data points are plotted in four figures depicted in Fig.~\ref{fig:Calibration}, which shows that the inverse of sensor output is increasing as the SWC increases, which follows a linear relation. So a linear regression model is used to fit all the points. In Table~\ref{tab: cal_mean_sd} the mean and standard deviation of three experimental replications has been shown. The mean and standard deviation are calculated using Equation~\ref{equ:8} and Equation~\ref{equ:9}. Accordingly, the goodness of fit and the Root Mean Square Error (RMSE) has been calculated. $R^2$ values, RMSE values and the calibration equation for three experiments are tabulated in Table~\ref{tab:rmse_r2_caleq}.
 
 \begin{equation}
\centering
    \bar x\ =\ \frac{\Sigma_{i=1}^n\ x_i}{n},
    \label{equ:8}
\end{equation}

\begin{equation}
    \sigma_1 = \sqrt\frac{\Sigma_{i=1}^n{(x_i\ -\ \bar x)^2}}{n}.
    \label{equ:9}
\end{equation}
Here  $x$ denotes the value of GWC, VWC and corresponding resistive and capacitive sensor output. $n$ is the number of repetition. $\bar x$ denoted the mean of the GWC, VWC and sensors output.
  \par The initial point in all the sub-figures in Fig.~\ref{fig:Calibration} resembles the data point for a dry soil sample. Fig.~\ref{fig:cal_gwc} and Fig.~\ref{fig:cal_vwc} is for capacitive sensor, where the initial point changes from experiment to experiment and Fig.~\ref{fig:res_gwc} and Fig.~\ref{fig: res_vwc} is for the resistive sensor response where it is constant for all three experiments.
\par
\begin{figure}[ht]
    \centering
    \subfigure[Capacitive Sensor]
    {
        \includegraphics[width = 4cm]{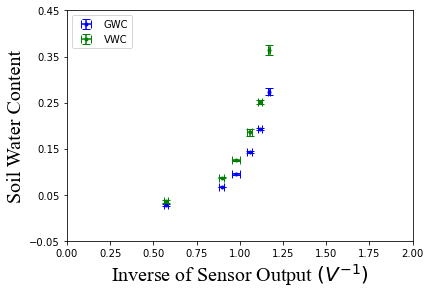}
        \label{fig:cap_err_swc}
    }
    \hfill
    \subfigure[Resistive Sensor]
    {
        \includegraphics[width = 4cm]{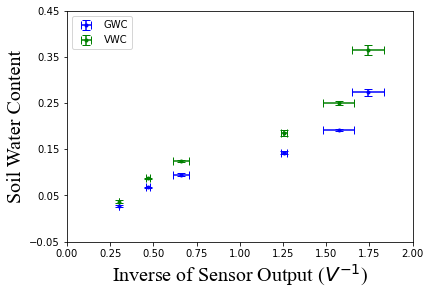}
        \label{fig:res_err_swc}
    }
    % \subfigure[Third caption]
    % {
    %     \includegraphics[width=1.0in]{imagefile2}
    %     \label{fig:third_sub}
    % }
    \caption{Response variation with mean and standard deviation}
    \label{fig:err_cal}
\end{figure}

 In Fig.~\ref{fig:err_cal} variation in the sensor response for the exact quantified soil water content measurement has been shown for capacitive and resistive sensors. The variance plot is made using the mean and standard deviation of the three experiments, data tabulated in Table~\ref{tab: cal_mean_sd}. The capacitive sensor response variation is less throughout than the resistive sensor. As the water content increases in the soil sample, the variation in the x-axis is increasing, i.e., the inverse of the resistive sensor output. In contrast, capacitive response variation is minimum for all the soil samples.

\begin{figure}[th]
    \centering
    \subfigure[Capacitive Sensor]
    {
        \includegraphics[width = 4cm]{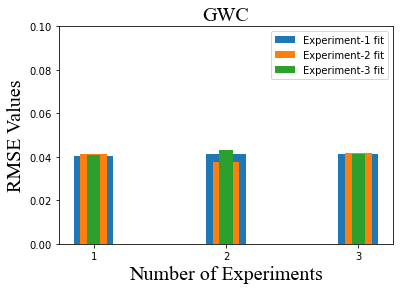}
        \label{fig:cap_rmse_gwc}
    }
    \hfill
    \subfigure[Capacitive Sensor]
    {
        \includegraphics[width = 4cm]{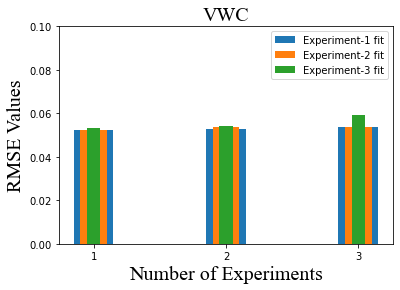}
        \label{fig:cap_rmse_vwc}
    }
    \subfigure[Resistive Sensor]
    {
        \includegraphics[width= 4cm]{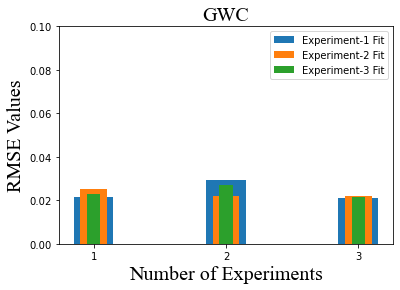}
        \label{fig:res_rmse_gwc}
    }
    \subfigure[Resistive Sensor]
    {
        \includegraphics[width = 4cm ]{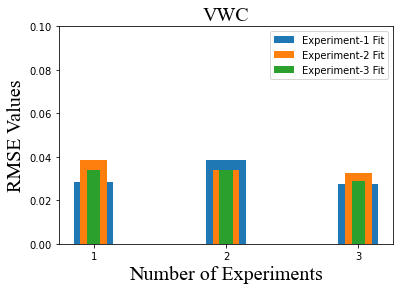}
        \label{fig:res_rmse_vwc}
    }
    \caption{Validation of the calibration equations with three experimental data}
    \label{fig:rmse}
\end{figure}

\begin{table*}[th]
\centering
\caption{Mean and Standard Deviation of three calibration experiment}
\begin{tabular}{|c|c|c|c|c|c|c|c|c|} 
\hline
\multirow{2}{*}{Container} & \multicolumn{4}{c|}{Mean}                                                                                                                                                              & \multicolumn{4}{c|}{Standard Deviation}                                                                                                                                                 \\ 
\cline{2-9}
                           & GWC    & VWC    & \begin{tabular}[c]{@{}l@{}}Capacitive Sensor\\Output ($V^{-1}$)\end{tabular} & \begin{tabular}[c]{@{}l@{}}Resistive Sensor\\Output ($V^{-1}$)\end{tabular} & GWC    & VWC    & \begin{tabular}[c]{@{}l@{}}Capacitive Sensor\\Output ($V^{-1}$)\end{tabular} & \begin{tabular}[c]{@{}l@{}}Resistive Sensor\\Output ($V^{-1}$)\end{tabular}  \\ 
\hline
1                          & 0.0276 & 0.0363 & 0.5718                                                                            & 0.303                                                                            & 0.0024 & 0.0033 & 0.0128                                                                            & 0                                                                                 \\ 
\hline
2                          & 0.0672 & 0.0873 & 0.8953                                                                            & 0.4682                                                                           & 0.0007 & 0.0011 & 0.0164                                                                            & 0.0108                                                                            \\ 
\hline
3                          & 0.0954 & 0.1247 & 0.9767                                                                            & 0.6595                                                                           & 0.0023 & 0.0011 & 0.0239                                                                            & 0.0477                                                                            \\ 
\hline
4                          & 0.1431 & 0.1854 & 1.0581                                                                            & 1.2565                                                                           & 0.0027 & 0.0072 & 0.0149                                                                            & 0.0166                                                                            \\ 
\hline
5                          & 0.1923 & 0.251  & 1.1169                                                                            & 1.5726                                                                           & 0.0023 & 0.0041 & 0.0138                                                                            & 0.0911                                                                            \\ 
\hline
6                          & 0.2737 & 0.3646 & 1.1705                                                                            & 1.7416                                                                           & 0.0079 & 0.0108 & 0.0068                                                                            & 0.0903                                                                            \\
\hline
\end{tabular}
\label{tab: cal_mean_sd}

\end{table*}
Validation of the calibration models is performed by using the calibration equations generated using the linear regression model fitted to the experimental data points. Three experiments have three calibration equations, each for GWC and VWC. Each equation is used to check the feasibility of predicting all the experimentally calculated values of the soil water content. The corresponding RMSE values are calculated and depicted in Fig.~\ref{fig:rmse}. The RMSE for estimating the GWC and VWC with a capacitive sensor varies from 0.0376 to 0.043 and 0.0523 to 0.0592, respectively. 
The RMSE ranges from 0.0211 to 0.0294 for resistive sensors predicting the GWC and 0.0277 to 0.0386 for resistive sensors predicting the VWC.

\begin{table*}
\centering
\caption{RMSE, $R^2$ and Calibration equation}
\begin{tabular}{|c|c|c|c|c|c|c|c|c|c|} 
\hline
\multirow{3}{*}{Sensor}                                                     & \multirow{3}{*}{\begin{tabular}[c]{@{}l@{}}No. of\\Experiment\end{tabular}} & \multicolumn{2}{c|}{RMSE}                   & \multicolumn{2}{c|}{R2 Score}               & \multicolumn{4}{c|}{Calibration Equation $y = mx + c$}  \\ 
\cline{3-10}
                                                                            &                                                                             & \multirow{2}{*}{GWC} & \multirow{2}{*}{VWC} & \multirow{2}{*}{GWC} & \multirow{2}{*}{VWC} & \multicolumn{2}{c|}{GWC} & \multicolumn{2}{c|}{VWC}       \\ 
\cline{7-10}
                                                                            &                                                                             &                      &                      &                      &                      & m      & c               & m      & c                     \\ 
\hline
\multirow{3}{*}{\begin{tabular}[c]{@{}l@{}}Resistive\\Sensor\end{tabular}}  & 1                                                                           & 0.0217               & 0.0284               & 0.9297               & 0.9258               & 0.135  & -0.0058         & 0.172  & -0.0066               \\ 
\cline{2-10}
                                                                            & 2                                                                           & 0.0222               & 0.0341               & 0.923                & 0.9018               & 0.1503 & -0.0119         & 0.2021 & -0.0188               \\ 
\cline{2-10}
                                                                            & 3                                                                           & 0.0214               & 0.0291               & 0.9287               & 0.9343               & 0.1446 & -0.0118         & 0.1956 & -0.0183               \\ 
\hline
\multirow{3}{*}{\begin{tabular}[c]{@{}l@{}}Capacitive\\Sensor\end{tabular}} & 1                                                                           & 0.0405               & 0.0523               & 0.754                & 0.749                & 0.4171 & -0.2315         & 0.5306 & -0.2935               \\ 
\cline{2-10}
                                                                            & 2                                                                           & 0.0376               & 0.0536               & 0.779                & 0.758                & 0.377  & -0.2016         & 0.5056 & -0.2728               \\ 
\cline{2-10}
                                                                            & 3                                                                           & 0.0412               & 0.0592               & 0.731                & 0.728                & 0.3948 & -0.211          & 0.5333 & -0.287                \\
\hline
\end{tabular}
\label{tab:rmse_r2_caleq}
\end{table*}

\subsection{Study of the Sensors Response}
\begin{figure}[!th]
    \centering
    \subfigure[]
    {
        \includegraphics[width = 4cm]{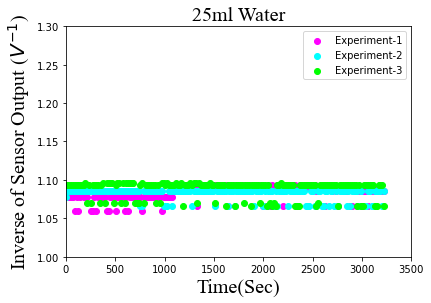}
        \label{fig:cap_25}
    }
    \hfill
    \subfigure[]
    {
        \includegraphics[width = 4cm]{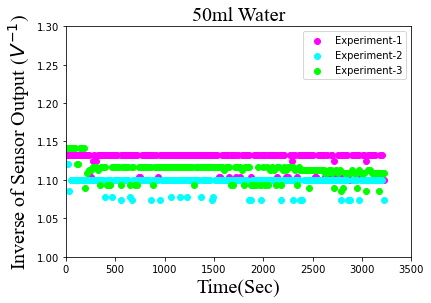}
        \label{fig:cap_50}
    }
    \subfigure[]
    {
        \includegraphics[width= 4cm]{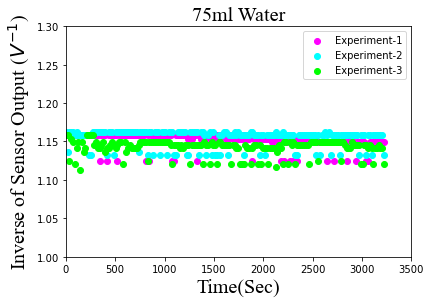}
        \label{fig:cap_75}
    }
    \subfigure[]
    {
        \includegraphics[width = 4cm ]{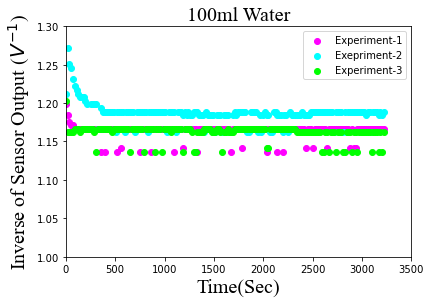}
        \label{fig:cap_100}
    }
    \caption{Capacitive sensor response for (a) 25ml, (b) 50ml, (c) 75ml and (d) 100ml water addition to a quantified amount of soil}
    \label{fig:cap_daytoday}
\end{figure}

\begin{figure}[th]
    \centering
    \subfigure[]
    {
        \includegraphics[width = 4cm]{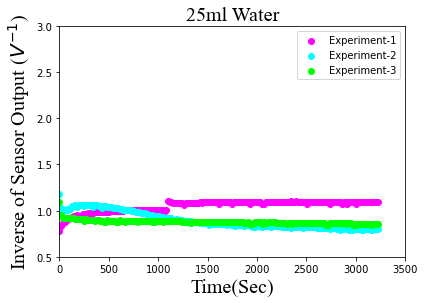}
        \label{fig:res_25}
    }
    \hfill
    \subfigure[]
    {
        \includegraphics[width = 4cm]{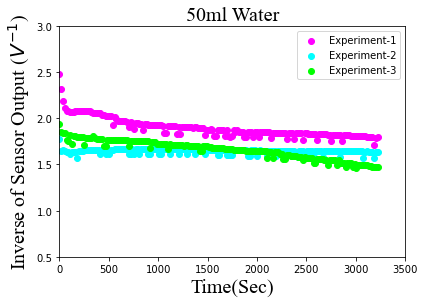}
        \label{fig:res_50}
    }
    \subfigure[]
    {
        \includegraphics[width= 4cm]{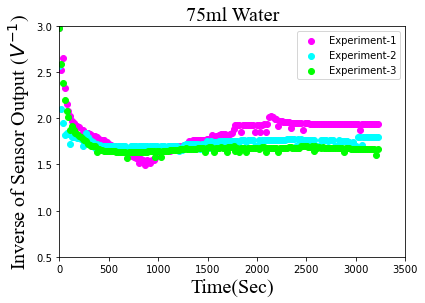}
        \label{fig:res_75}
    }
    \subfigure[]
    {
        \includegraphics[width = 4cm ]{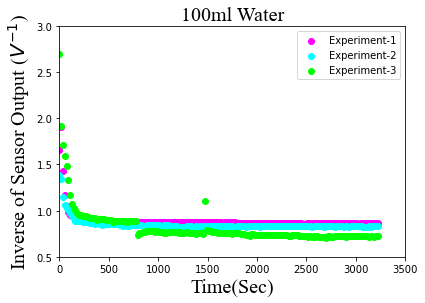}
        \label{fig:res_100}
    }
\caption{Resistive sensor response for (a) 25ml, (b) 50ml, (c) 75ml and (d) 100ml water addition to a quantified amount of soil}
\label{fig:res_daytoday}
\end{figure}

Understanding the response of the soil moisture sensor is very important to get an insight into the SWC from the sensor response. The comparative study experiment is performed in the laboratory in a controlled (room temperature) environment for one hour. The experimental setup includes four containers of the same size and volume, soil, water, and sensor system. 375gm soil in each container mixed with water and the sensor probes are inserted into the soil to read the moisture data. Thingspeak is used for storing sensor data in two different channels for respective sensors. Three repetitions of the experiment have been made to account for the sensor's reliability. Fig.~\ref{fig:cap_daytoday} and Fig.~\ref{fig:res_daytoday} show the day to day responses of the capacitive and resistive sensor with time.
\begin{figure}
    \centering
    \includegraphics[width = 8cm]{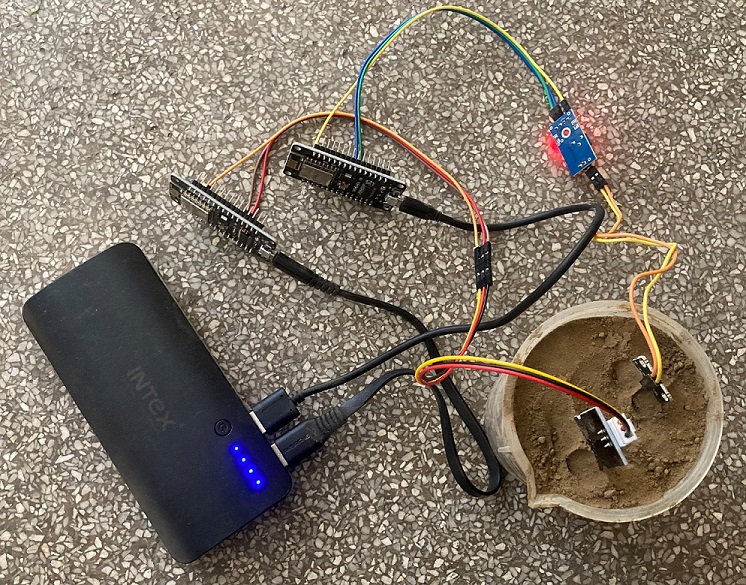}
    \caption{Data collection for the response of capacitive and resistive sensor}
    \label{fig:comparative_data_collection}
\end{figure}

\begin{table}[th]

\caption{Mean and Standard Deviation of Comparative Study}
\centering

\begin{tabular}{|c|c|c|c|c|c|c|} 
\hline
\multirow{2}{*}{Sensor}                                                     & \multirow{2}{*}{\begin{tabular}[c]{@{}l@{}}Statistical\\Measure\end{tabular}} & \multirow{2}{*}{Day} & \multicolumn{4}{c|}{Water content}  \\ 
\cline{4-7}
                                                                            &                                                                               &                             & 25ml   & 50ml   & 75ml   & 100ml    \\ 
\hline
\multirow{6}{*}{\begin{tabular}[c]{@{}c@{}}Capacitive\\Sensor\end{tabular}} & \multirow{3}{*}{\begin{tabular}[c]{@{}c@{}}Mean\\($\bar x$)\end{tabular}}                                                         & 1                           & 1.0829 & 1.1272 & 1.1511 & 1.1638   \\ 
\cline{3-7}
                                                                            &                                                                               & 2                           & 1.0813 & 1.097 & 1.1547 & 1.1867   \\ 
\cline{3-7}
                                                                            &                                                                               & 3                           & 1.088  & 1.1114 & 1.1417 & 1.1617   \\ 
\cline{2-7}
                                                                            & \multirow{3}{*}{\begin{tabular}[c]{@{}c@{}}Standard\\Deviation\\($\sigma$)\end{tabular}}  & 1                           & 0.0102 & 0.011  & 0.0101 & 0.0094   \\ 
\cline{3-7}
                                                                            &                                                                               & 2                           & 0.0074 & 0.0092 & 0.0108 & 0.0146   \\ 
\cline{3-7}
                                                                            &                                                                               & 3                           & 0.0103 & 0.0114 & 0.0102 & 0.0098   \\ 
\hline
\multirow{6}{*}{\begin{tabular}[c|]{@{}l@{}}Resistive\\Sensor\end{tabular}}  & \multirow{3}{*}{\begin{tabular}[c|]{@{}c@{}}Mean\\($\bar x$)\end{tabular}}                                                         & 1                           & 1.0512 & 1.8953 & 1.8448 & 0.8883   \\ 
\cline{3-7}
                                                                            &                                                                               & 2                           & 0.8946 & 1.6443 & 1.7432 & 0.8469   \\ 
\cline{3-7}
                                                                            &                                                                               & 3                           & 0.8777 & 1.6607 & 1.6969 & 0.8172   \\ 
\cline{2-7}
                                                                            & \multirow{3}{*}{\begin{tabular}[c|]{@{}c@{}}Standard\\Deviation\\($\sigma$)\end{tabular}}  & 1                           & 0.0629 & 0.1089 & 0.1836 & 0.1087   \\ 
\cline{3-7}
                                                                            &                                                                               & 2                           & 0.0922 & 0.0261 & 0.0801 & 0.0678   \\ 
\cline{3-7}
                                                                            &                                                                               & 3                           & 0.0238 & 0.1001 & 0.147 & 0.2157    \\
\hline
\end{tabular}
\label{tab: mean_sd_comp}
\end{table}
The mean and standard deviation of the response data of each sensor response for each quantified amount of soil and water are calculated using Equation~\ref{equ:8} and Equation~\ref{equ:10}.
\begin{equation}
    \sigma_2 = \sqrt\frac{\Sigma_{i=1}^n{(x_i\ -\ \bar x)^2}}{n-1}.
    \label{equ:10}
\end{equation}

Here $x$ denotes the amount of water mixed with the soil and $n$ denotes the number of data points considered for each water content, $\bar x$ is the mean, and $\sigma_2$ is the standard deviation. The moisture measurements have a total of 180 data points for each quantity of water each day. The calculated mean and standard deviation are given in Table~\ref{tab: mean_sd_comp}.

\par
\begin{figure}[th]
    \centering
    \subfigure[]
    {
        \includegraphics[width = 4cm]{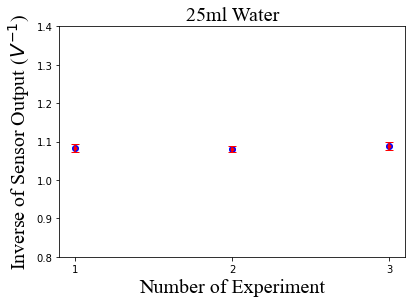}
        \label{fig:cap_25_err}
    }
    \hfill
    \subfigure[]
    {
        \includegraphics[width = 4cm]{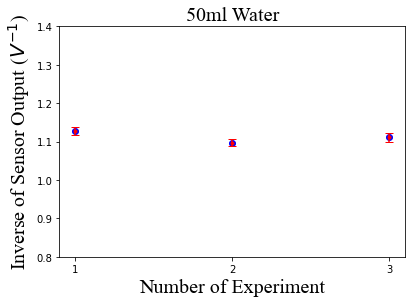}
        \label{fig:cap_50_err}
    }
    \subfigure[]
    {
        \includegraphics[width= 4cm]{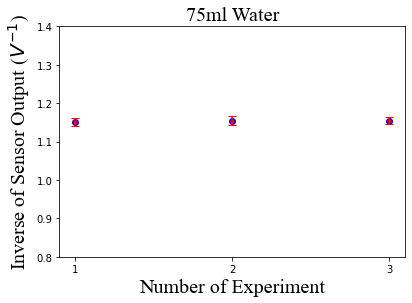}
        \label{fig:cap_75_err}
    }
    \subfigure[]
    {
        \includegraphics[width = 4cm ]{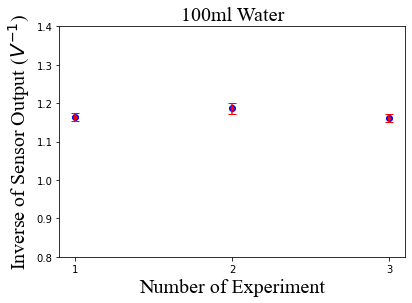}
        \label{fig:cap_100_err}
    }
    \caption{Variations in Capacitive sensor response shown using the mean and standard deviation of three experiment}
    \label{fig:cap_err}
\end{figure}

\begin{figure}[th]
    \centering
    \subfigure[]
    {
        \includegraphics[width = 4cm]{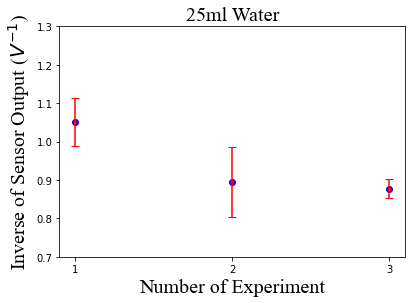}
        \label{fig:res_25_err}
    }
    \hfill
    \subfigure[]
    {
        \includegraphics[width = 4cm]{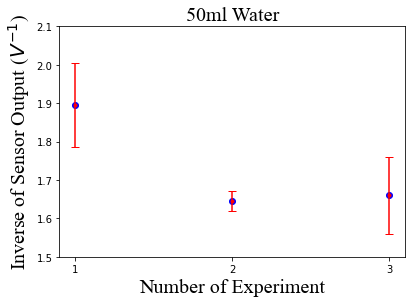}
        \label{fig:res_50_err}
    }
    \subfigure[]
    {
        \includegraphics[width= 4cm]{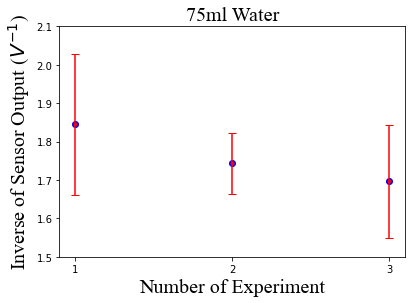}
        \label{fig:res_75_err}
    }
    \subfigure[]
    {
        \includegraphics[width = 4cm ]{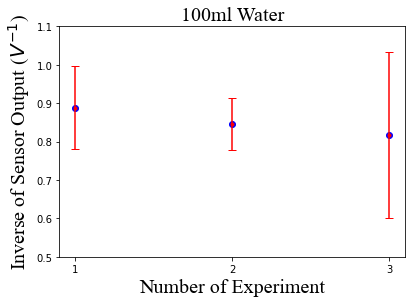}
        \label{fig:res_100_err}
    }
    \caption{Variation in Resistive sensor response using the mean and standard deviation of three experiment}
    \label{fig:res_err}
\end{figure}

 Fig.~\ref{fig:cap_err} and Fig.~\ref{fig:res_err} shows the resistive and capacitive sensor response variation from the mean with standard deviation in day to day experiments. The span of the y-axis in Fig.~\ref{fig:cap_err} and Fig.~\ref{fig:res_err} are intentionally kept same to distinguish the variation in y-axis i.e.,resistive and capacitive sensor response.

\section{Discussion and Conclusion}
The comparison based on the calibration and response of two low-cost soil moisture sensors is presented in this manuscript. The calibration equations for GWC and VWC are shown in Table~\ref{tab:rmse_r2_caleq}. The effect of repetitions on the calibration is shown using the variance plot in Fig.~\ref{fig:err_cal}. The RMSE and $R^2$ value are calculated for the corresponding calibration equation. Validation of the calibration equations is investigated by taking each equation for predicting the GWC and VWC of each experiment. The response for each amount of water addition is plotted against time (1 hour), and the variation plot for the repetitive experiment has been shown in Fig.~\ref{fig:cap_daytoday} and Fig.~\ref{fig:res_daytoday}.
The calibration result is analyzed based on the $R^2$ value and the  RMSE value, based on which resistive sensor performance is better than the capacitive sensor. However, there are some contradictions observed from the calibration experiment, i.e., the dry soil resistive sensor could not detect the change in moisture content from one experiment to another, whereas the capacitive sensor detected. It can be said that the resistive sensor is less sensitive than the capacitive sensor.
\par In the response study the observation drawn from Fig.~\ref{fig:res_25} and Fig.~\ref{fig:res_100} is that the resistive sensor response for 25ml water addition and 100ml water addition is almost same. This result can be related to the calibration result that the variation of the resistive sensor increases with the increase in SWC. So, the abnormality in resistive sensor response can happen when the soil reaches its saturation level. In the case of the capacitive sensor, no such abnormality was observed. From the Fig.~\ref{fig:cap_err} and Fig.~\ref{fig:res_err}, it can be noticed that the output variation is larger in resistive sensor response than the capacitive sensor response.
\par Comparing the resistive and capacitive sensors reveals that the resistive sensor is unreliable for measuring soil moisture since its response varies from one experiment to another. Another discovery from the response analysis is that the soil-exposed resistive sensor probes are sensitive to corrosion, so after 5-6 short-duration measurements, the sensor loses its functionality. For the capacitive sensor, however, the probes are coated with a polymer that resists electrolysis and oxidation. Consequently, the capacitive sensor is suitable for long-duration measurements, whereas the resistive sensor is ideal for instant measurement.
\par Thus, this type of analysis is required before using low-cost moisture sensors to find their suitability with the user application context. Calibration is essential before getting started with moisture sensing or integrating moisture sensors in any automated moisture monitoring platform.
\par This paper discusses the calibration and the sensor response for different moisture levels for two low-cost soil moisture sensors. The inference from the discussion is helpful for large-scale soil moisture measurement, usually for noncontact measurement, where the spatial resolution is significantly less, say 10 meters to a few kilometers. This work is useful in validating and visualizing the ground truth information on SM for noncontact sensing technologies. Optical remote sensing and other noncontact methods are used for numerous agricultural applications, and estimation of soil water content is one of them. Though it gives us some idea regarding the amount of surface soil moisture locally or regionally, it fails to provide the point values in VWC. 
\par
\begin{figure}[th]
    \centering
    \includegraphics[width=8cm]{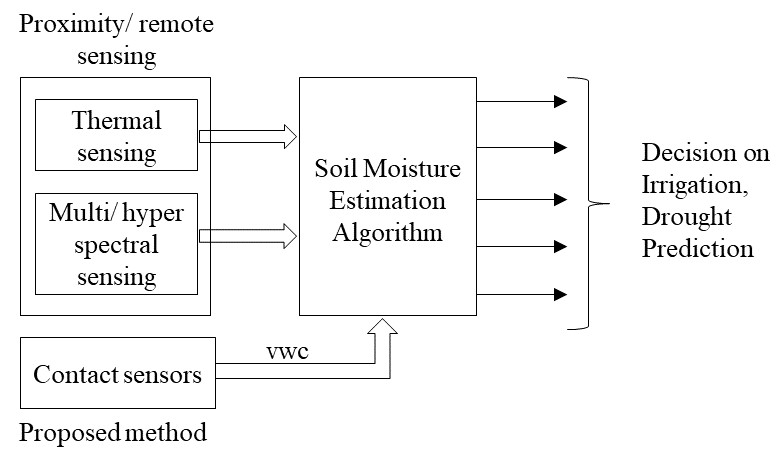}
    \caption{Overview on Soil Moisture Estimation and Decision Making using Contact and Noncontact sensors}
    \label{fig:overview}
\end{figure}

 The results and work analysis are beneficial before starting with an automatic moisture monitoring platform. After selecting an appropriate sensor based on the measurement criterion and need, the data acquisition system mentioned in this paper can be used to evaluate and validate the SWC. Further, this setup can be used with noncontact type sensors to estimate the soil moisture locally or regionally. The noncontact type of soil moisture measurement, mainly the optical sensors, is used for numerous agricultural applications; estimation of soil water content is one. Though these sensors have a high spatial resolution, they provide an idea of the surface soil moisture locally or regionally but fail to give the point values in VWC. Another advantage of these sensors is that they can be used to obtain both surface and root zone soil moisture information. Still, the proximity or optical remote sensing technique fails to do the latter. Fig.~\ref{fig:overview} gives an overview of the applicability of the presented work to soil moisture estimation along with the usage of noncontact sensors. 

\bibliographystyle{IEEEtran}
\bibliography{reference.bib}

\end{document}